# A longitudinal geospatial multimodal dataset of post-discharge frailty, physiology, mobility, and neighborhoods

Ali Abedi [1,2] *, Charlene H. Chu [1,2], and Shehroz S. Khan [3]

*Correspondence: ali.abedi@uhn.ca
[1] Lawrence Bloomberg Faculty of Nursing, University of Toronto, Toronto, Canada
[2] KITE Research Institute, Toronto Rehabilitation Institute, University Health Network, Toronto, Canada
[3] College of Engineering and Technology, American University of the Middle East, Kuwait

## Abstract
Frailty in older adults is associated with increased vulnerability to functional decline, reduced mobility, social isolation, and challenges during the transition from hospital to community living. These factors are associated with rehospitalization and may adversely influence recovery. Neighborhood environments can further shape recovery trajectories by affecting mobility opportunities, social engagement, and access to community resources. Multimodal sensing technologies combined with data-driven analytical approaches offer the potential to continuously monitor these multidimensional factors in real-world settings. This Data Descriptor presents GEOFRAIL, a longitudinal geospatial multimodal dataset collected from community-dwelling frail older adults following hospital discharge. The dataset is organized into interconnected tables capturing participant demographics, features derived from multimodal sensors, biweekly clinical assessments of frailty, physical function, and social isolation, and temporal location records linked to neighborhood amenities, crime rates, and census-based socioeconomic indicators. Data were collected over an eight-week post-discharge period using standardized pipelines with privacy-preserving spatial aggregation. Technical validation demonstrates internal consistency across geospatial, sensor-derived, and clinical measures and reports baseline performance of machine learning models for characterizing recovery trajectories.

## Background & Summary
Frailty is a major determinant of vulnerability among older adults and is closely linked to hip and other lower limb fractures, which both contribute to and result from frailty[1,2]. Reduced muscle strength, impaired balance, and decreased physiological reserve increase fracture risk, while fracture-related immobility, hospitalization, and surgical stress can accelerate frailty progression[3,4]. Following hip and other lower limb fractures, many older adults experience prolonged mobility limitations[5], loss of independence[6], and increased susceptibility to functional decline[2], social isolation[7], and rehospitalization[8,9] during the transition from hospital to community living.

In the period of at-home recovery following hospital discharge for hip and other lower limb fractures, mobility and social engagement are central to recovery[1,2,7,10]. The ability of older adults to leave the home and age in place[11,12] is strongly influenced by the neighborhood environments in which they live and move[13–15]. Access to nearby amenities such as parks, community centers, libraries, and cafes can encourage going out, support regular physical mobility, and facilitate social interaction[16,17]. In contrast, neighborhoods with limited amenities, lower socioeconomic resources, or higher crime rates may discourage outdoor activity and social participation due to safety concerns or lack of accessible destinations[16–20]. Such constraints can reinforce sedentary behavior and social isolation, accelerating functional decline and negatively influencing recovery trajectories. Social isolation frequently co-occurs with frailty, particularly among older adults living alone, and is closely linked to poorer physical and mental health outcomes. Neighborhood-level socioeconomic conditions and perceived safety can either mitigate or exacerbate isolation by shaping opportunities for social engagement and community participation[16–20]. Despite the importance of these contextual factors, post-discharge care typically relies on episodic clinic-based assessments and rarely incorporates continuous monitoring of daily behavior or neighborhood context[21]. As a result, gradual but clinically meaningful changes in mobility, social interaction, and recovery often remain undetected until they contribute to significant health setbacks or rehospitalization[22].

Advances in wearable sensing technologies, the Internet of Things, and data-driven analytical methods provide new opportunities to address these gaps[23–25]. Multimodal sensors enable continuous, passive measurement of physiological signals, mobility patterns, and daily behaviors in real-world settings[23–25]. When integrated with geospatial and socioeconomic neighborhood data[26–28], these approaches can enable a more comprehensive understanding of how individual frailty interacts with environmental context to influence post-discharge recovery. To support research at this intersection, this Data Descriptor presents the GEOspatial FRAILty and Recovery in Independent Living (GEOFRAIL) dataset. It integrates multimodal sensor data, clinical assessments of frailty and social isolation, and neighborhood-level amenity, crime, and census indicators from frail older adults after hospital discharge for lower limb fracture. This resource enables the investigation of the joint influence of place-based factors, daily behaviors, and frailty on recovery trajectories.

Existing datasets and studies in older adult populations have largely focused on either individual-level sensing or neighborhood-level contextual characteristics, but rarely on their integration[23–28]. Multiple datasets and studies have leveraged wearable or in-home sensors to capture longitudinal physiological signals, physical activity, gait, sleep, and daily behavior in community-dwelling or post-discharge older adults, supporting applications in frailty assessment and recovery monitoring[22,29]. Recent publicly available datasets similarly emphasize multimodal sensing in aging populations, including in-home and wearable monitoring of activity and sleep, but do not incorporate neighborhood-level socioeconomic or environmental context[24,30]. In parallel, a separate literature examines neighborhood determinants of health using census or geographic information systems data without longitudinal individual-level sensing[11–14]. Collectively, this literature underscores the absence of resources that integrate longitudinal multimodal sensor data with neighborhood-level socioeconomic and environmental characteristics to study post-discharge recovery and frailty in older adults, motivating the development of the GEOFRAIL dataset to support research at this intersection.

## Methods

This section describes the collection and integration of data in the GEOFRAIL dataset. Longitudinal multimodal sensor data and repeated clinical assessments were collected during at-home recovery after hospital discharge. These data were processed and linked to neighborhood-level socioeconomic, environmental, and safety characteristics using deidentified geospatial identifiers.

### *Participants*

Frail older adults who experienced a lower limb fracture were recruited from the Toronto Rehabilitation Institute, University Health Network, Toronto, Canada. All data collection was conducted within the Greater Toronto Area, Canada. The study was approved by the University Health Network Research Ethics Board (study ID 20-5113.10), and written informed consent was obtained from all participants, permitting public release of de-identified data.

Inclusion criteria required participants to be 60 years of age or older and to have undergone surgical repair for a hip, femur, or pelvis fracture, or to have received a hip replacement. Participants were also required to have a home Wi-Fi connection and no cognitive impairment. Recruitment and enrollment occurred prior to hospital discharge, allowing data collection to begin within the first few days following discharge and continue for eight weeks.

### *Multimodal sensor and clinical data collection*

Multimodal AI-based Sensor platform for Older iNdividuals (MAISON)[31] is a remote patient monitoring platform designed for longitudinal data collection in community-dwelling older adults. It integrates smartwatches, smartphones, and external wearable and non-wearable sensors to enable continuous and automatic acquisition of physiological, activity, sleep, location, and movement data[31]. Data collection parameters, including sampling frequency and duration, are configurable by study. Location tracking is activated only when participants leave the home using geofencing to reduce battery burden. External commercial sensors can be incorporated when manufacturer-provided interfaces allow secure data access, supporting unified management of multimodal data.

**Multimodal sensor data collection.** For this study, MAISON deployment included the following devices and data modalities:

- a Google Pixel Watch 2 Wear OS smartwatch collecting acceleration every second, heart rate every 30 minutes, step count continuously, and global positioning system (GPS) location data at one-minute intervals when participants were outside the home geofence, defined as a circular buffer with a 50 m radius centered on the participant's home location;
- a Motorola Moto G54 Android smartphone used to administer clinical questionnaires;
- a Proteus M5 motion sensor installed in the living room to detect in-home movement events; and
- a Withings Sleep under-mattress sleep-tracking mat to record sleep patterns, including total sleep time, deep, light, and rapid eye movement sleep duration, snoring duration, sleep and wake latency, heart rate during sleep, and wake-up count.

Data collection and transfer to the MAISON private cloud on Google Cloud Platform[31] were continuous and automatic, requiring no participant effort beyond routine device charging and wearing the smartwatch. A trained research assistant installed the in-home sensors at the participant's residence for an eight-week monitoring period and removed them upon completion of data collection.

**Frailty clinical assessment.** In accordance with Fried's frailty phenotype[32], clinical and physical frailty assessments were administered by a trained research assistant at the sensor installation and sensor removal visits, including:
- Grip strength: Measured using a validated handheld dynamometer (CAMRY EH101)[33]. Participants were instructed to grip the device as firmly as possible for 5 seconds with each hand, and the recorded values were documented.
- Body weight: Measured using a validated digital scale (Withings Body). Participants stood on the scale for approximately 5 seconds while the displayed weight was recorded.
- Physical function: Assessed using the Timed Up and Go test (TUG)[34].
- Physical activity: Assessed using the 9-item Rapid Assessment of Physical Activity (RAPA) questionnaire[35].
- Frailty status: Evaluated using the Clinical Frailty Scale Health Questionnaire (CFS-HQ)[36], a 7-domain instrument used to inform assignment of the 9-point Clinical Frailty Scale score.

**Physical functioning and social interaction assessment.** Validated clinical assessments were administered every two weeks by a trained research assistant via Microsoft Teams video calls with participants at home. These assessments provided reference measures of social interaction and physical function and included the 6-item Social Isolation Scale (SIS)[37], the 12-item Oxford Hip Score (OHS)[38], the 12-item Oxford Knee Score (OKS)[39], the TUG test[34], and the 30-second chair stand test[40]. Detailed questionnaire content, including the full list of items and test administration instructions, is provided in Supplementary Information 1. Figure 1 illustrates the devices used for data collection in the MAISON platform.

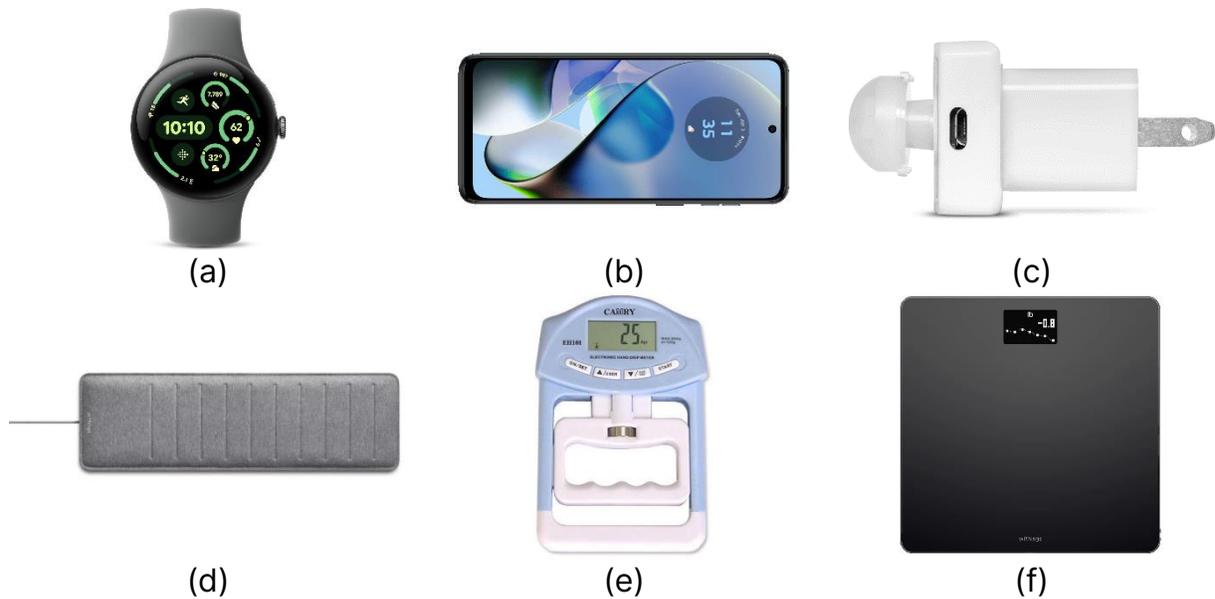

Figure 1. Data collection devices used in the MAISON platform: (a) Google Pixel Watch 2 Wear OS smartwatch, (b) Motorola Moto G54 Android smartphone, (c) Proteus M5 motion sensor, (d) Withings Sleep under-mattress sleep-tracking mat, (e) CAMRY EH101 handgrip dynamometer, and (f) Withings Body digital scale.

### *Neighborhood-level data integration*

GPS location data collected via the smartwatch in the MAISON platform includes timestamps and latitude and longitude (geographic coordinates) for locations participants visited while outside the home geofence. These data were used to derive three neighborhood-level contextual measures, including neighborhood amenities, crime rates, and census-based socioeconomic indicators (refer to Figure 2).

**Amenities data.** Latitude and longitude coordinates for each unique participant location were used as the center point to quantify neighborhood amenities within a 1 km radius. For each coordinate pair, the Google Places Nearby Search API[41] was queried to count the number of nearby points of interest in predefined categories, including parks, libraries, and restaurants or cafes, as well as keyword-based searches for community centers and places of worship. API pagination was handled using returned page tokens to ensure all results within the search radius were counted.

Latitude and longitude coordinates were reverse geocoded using the Google Geocoding API[42] to obtain formatted address information for each recorded location. Canadian six-character postal codes were then extracted from the returned address strings and assigned to corresponding location records for linkage with neighborhood-level data.

**Crime rates data.** Crime rates were derived by linking participant location records to publicly available Neighborhood Crime Rates Open Data[43] provided by the Toronto Police Service. For each unique Canadian postal code associated with participant locations, the corresponding Forward Sortation Area (FSA)[44] was used to identify the associated neighborhood. Neighborhood names were matched to official Toronto Police Service crime records using approximate string matching to account for naming inconsistencies. Annual neighborhood-level crime rate indicators, including assault, robbery, theft, break and enter, auto theft, and related offenses, were extracted and assigned to participant location records.

**Census-based socioeconomic data.** Census-based socioeconomic characteristics were derived by linking participant location records to publicly available data from Statistics Canada[45]. Canadian postal codes associated with participant locations were first mapped to dissemination areas[46], the smallest standard geographic unit used by Statistics Canada, using the Postal Code Conversion File (PCCF)[47]. These dissemination areas were then linked to the Statistics Canada Census Profile[45] to extract comprehensive neighborhood-level sociodemographic information, including population composition, income, education, housing, and employment characteristics. Additional sociodemographic dimensions encompassed educational attainment, immigration and citizenship status, visible minority and ethnic origin categories, language knowledge, household composition, housing tenure and suitability, and mobility status. Variables captured population counts and sex-stratified distributions using total counts, counts for men and women, along with corresponding proportions or rates[45].

Together, these variables provided a detailed neighborhood-level context encompassing amenity proximity, crime rates, and socioeconomic and demographic characteristics. This integration enables examination of how local population structure, economic resources, and social composition relate to mobility patterns, social interaction, and recovery trajectories among community-dwelling older adults.

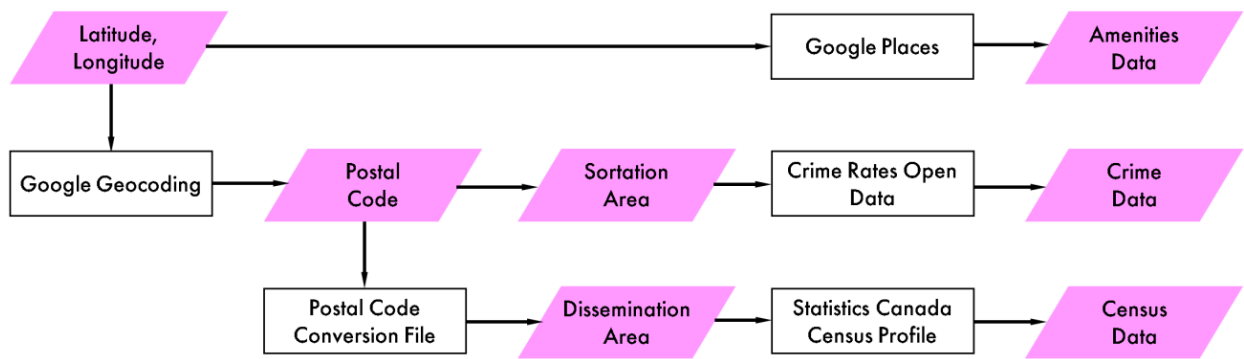

Figure 2. Designed pipeline linking GPS location data to neighborhood-level information, including amenities, crime rates, and socioeconomic characteristics of locations participants visited or stayed in.

## Data Records

The GEOFRAIL dataset (version v1) is hosted on Zenodo[48]. The dataset consists of the following interconnected tables linked by shared primary and foreign keys, which are denoted by underlining below.

To protect participant privacy, the dataset has been de-identified. Raw latitude, longitude, and postal code information are not shared. Locations where participants resided or visited are represented exclusively using coded location identifiers (L0001–L4096). In addition, small random perturbations (3–5%) were applied to non-zero amenities, crime rate, and census-derived variables to reduce the risk of re-identification while preserving aggregate distributions. The dates on which participant data were collected were also temporally perturbed to further limit the risk of re-identification.

1) **Demographics.** The first table (18 × 11) contains participant-level demographic and study information, where 18 represents the number of participants and 11 represents the number of variables, including:
   - participant ID (1 variable);
   - monitoring period start and end dates (2 variables);
   - demographic characteristics: age, sex, ethnicity, education level, employment status, relationship status, and fracture type (7 variables); and
   - location ID of participant's home (1 variable).

2) **Sensor features.** The second table (18 × 56 × 48) contains features extracted on a daily basis from multiple continuous multimodal sensor streams collected from participants, as described in Abedi et al.[24]. Each feature represents a specific characteristic of a sensor modality on a day; for example, one step-derived feature reports the total number of steps per day. In total, 46 sensor-derived features are provided per participant per day, with 56 days of monitoring per participant. Detailed information on the features extracted from different data modalities is provided in Supplementary Information 2.
   - participant ID (1 variable);
   - timestamp date (1 variable);
   - acceleration (16 variables); heart rate (6 variables); motion (5 variables); location (3 variables); sleep (11 variables); and step features (5 variables).

3) **Clinical.** The third table (18 × 4 × 63) contains clinical questionnaire responses and physical performance test results collected biweekly or every 8 weeks.
   - participant ID (1 variable);
   - timestamp bi-weekly or every 8 weeks date (1 variable);
   - SIS (6 variables); OHS (12 variables); OKS (12 variables); TUG test (1 variable); 30-second Chair Stand test (1 variable); RAPA (9 variables); CFS-HQ (16 variables); grip strength (2 variables); and body weight and height (2 variables)

4) **Temporal location.** The fourth table (18,843 × 3) contains deidentified temporal location identifiers representing locations where participants stayed or visited.
   - participant ID (1 variable);
   - timestamp second level (1 variable); and
   - location ID of locations participants stayed in or visited (1 variable).

5) **Amenities.** The fifth table (2,881 × 6) contains deidentified unique location IDs linked to amenity proximity data, including counts of selected amenities within a 1 km radius of locations where participants stayed or visited. These 2,881 unique location IDs account for 18,843 temporal location records, reflecting repeated visits or stays at the same locations across participants and time.
   - location ID (1 variable);
   - number of community centers; libraries; parks; food establishments; and places of worship (5 variables).

6) **Crime rate.** The sixth table (2,881 × 10) contains deidentified unique location IDs linked to crime rates in the neighborhoods where participants stayed or visited.
   - location ID (1 variable);
   - assault; auto theft; bike theft; break and enter; homicide; robbery; shooting; theft from motor vehicle; and high value theft rate (9 variables).

7) **Census.** The seventh table (2,881 × 1,624 × 6) contains deidentified unique location IDs linked to 1,624 census variables describing neighborhood level characteristics of the areas where participants stayed or visited, reported across six population measures, total count, men count, women count, total rate, men rate, and women rate. Detailed information on the census variables is provided in Supplementary Information 3.
   - location ID (1 variable);
   - census data (1,624 × 6 variables).

## Data Overview

GEOFRAIL comprises data collected from 18 participants, each monitored for eight weeks, resulting in 1,008 days of continuous multimodal sensor data alongside repeated clinical questionnaire assessments and physical performance tests, all linked to deidentified neighborhood-level contextual data. Data were collected between March 2022 and October 2025, with participants enrolled across four calendar years. The cohort had a mean age of 76.5 years (SD 8.9) and included 14 females and 4 males, with participants primarily identifying as Caucasian (N = 15) and additional representation from Black (N = 2) and South Asian (N = 1) backgrounds.

Table 1 provides an overview of the scale and distribution of selected variables in the dataset, spanning participant-level measures and location-level contextual characteristics. The table summarizes central tendency, variability, range, and distributional shape for representative variables drawn from multiple data sources, including demographic attributes, mobility-derived measures, physical performance outcomes, social measures, neighborhood amenities, crime rates, and socioeconomic indicators. Participant ages span a broad range, while mobility-related measures such as the number of GPS location data points and time spent outside the home show substantial dispersion and right skew, reflecting heterogeneity in movement patterns across individuals. Changes in grip strength and social isolation scores are centered closer to zero with wider spread, indicating both increases and decreases over the monitoring period. In contrast, neighborhood-level variables, including amenity counts, crime rates, and socioeconomic indicators, exhibit larger ranges and stronger positive skewness, consistent with uneven spatial distributions across locations.

## Technical Validation

This section evaluates the suitability of the GEOFRAIL dataset for statistical analyses examining relationships among multimodal sensor data, neighborhood-level contextual variables, and clinical outcomes, as well as for machine learning–based approaches that leverage sensor and neighborhood data to detect and predict clinical outcomes.

The association was assessed using a proportional-odds ordinal logistic regression. The outcome was treated as an ordered categorical variable, and the predictor was standardized so that estimates correspond to a one standard-deviation increase. Results are reported as an odds ratio with 95 percent confidence intervals and a $p$-value. Table 2 presents the results of association analyses examining illustrative multimodal sensor-derived features and GPS-derived neighborhood-level variables in relation to representative clinical outcomes, including SIS, OHS, and RAPA.

Table 1. Summary statistics, mean, standard deviation, minimum, maximum, and skewness, for illustrative variables computed across participants (superscript $p$) or unique participant-visited locations (superscript $ul$).

| Variable | Mean | SD | Min | Max | Skew |
|---|---|---|---|---|---|
| Age $^p$ | 76.5 | 8.64 | 60.00 | 94.00 | 0.13 |
| Number of GPS location data points $^p$ | 1,046.83 | 672.60 | 0 | 2,262 | 0.021 |
| Duration (hours) spent outside $^p$ | 96.77 | 115.18 | 0 | 483.73 | 2.08 |
| Grip strength (kg) of left hand (initial value) $^p$ | 15.73 | 4.14 | 7.80 | 23.50 | -0.06 |
| Grip strength (kg) of left hand (changes) $^p$ | 3.23 | 5.17 | -0.50 | 16.30 | 1.65 |
| Social Isolation Scale (initial value) $^p$ | 23.00 | 4.94 | 15.00 | 29.00 | -0.41 |
| Social Isolation Scale (changes) $^p$ | 0.11 | 2.96 | -6.00 | 5.00 | -0.17 |
| Number of community centers $^{ul}$ | 2.16 | 2.03 | 0 | 11.00 | 1.05 |
| Number of food establishments $^{ul}$ | 8.28 | 4.81 | 0 | 20.00 | 0.34 |
| Assault rate (per 100,000 people per year) $^{ul}$ | 1,204.30 | 857.17 | 183.97 | 2616.92 | 0.79 |
| Homicide rate (per 100,000 people per year) $^{ul}$ | 3.47 | 4.27 | 0 | 23.16 | 0.87 |
| Average after-tax income $^{ul}$ | 62,871.18 | 34,732.42 | 13,850.0 | 444,000.00 | 3.61 |
| Visible minority population $^{ul}$ | 1,076.81 | 1,343.52 | 150.00 | 11,190.0 | 4.14 |

Table 2. Association analysis between illustrative multimodal sensor–derived features and GPS-derived neighborhood-level variables in relation to representative clinical outcomes, including the Social Isolation Scale (SIS), Oxford Hip Score (OHS), and Rapid Assessment of Physical Activity (RAPA). Sensor-driven associations were calculated across days of data collection and participants (superscript $d$), whereas neighborhood-level associations were calculated across the locations visited by participants (superscript $vl$). †Each cell reports odds ratio (95% confidence intervals) and $p$-value.

| Variable | SIS | OHS | RAPA |
|---|---|---|---|
| Acceleration signal standard deviation $^d$ | 2.132 (1.902–2.390), < 0.001† | 0.870 (0.785–0.964), 0.0081 | 1.850 (1.569–2.182), < 0.001 |
| Heart rate minimum $^d$ | 0.461 (0.399–0.532), < 0.001 | 1.167 (1.061–1.284), 0.0015 | 0.513 (0.423–0.622), < 0.001 |
| Sleep onset latency $^d$ | 0.713 (0.642–0.792), < 0.001 | 0.743 (0.665–0.829), < 0.001 | 0.843 (0.726–0.979), 0.0248 |
| Sleep mean heart rate $^d$ | 0.594 (0.530–0.667), < 0.001 | 1.130 (1.021–1.250), 0.0182 | 0.551 (0.466–0.650), < 0.001 |
| Step count $^d$ | 1.333 (1.198–1.483), < 0.001 | 0.895 (0.815–0.984), 0.0218 | 1.273 (1.103–1.469), < 0.001 |
| Motion count $^d$ | 1.230 (1.124–1.347), < 0.001 | 1.124 (1.028–1.229), 0.0103 | 1.094 (0.957–1.251), 0.1865 |
| Number of GPS location data points $^d$ | 1.323 (1.193–1.468), < 0.001 | 1.083 (0.982–1.193), 0.1089 | 1.033 (0.895–1.191), 0.6603 |
| Number of community centers $^{vl}$ | 1.329 (1.294–1.365), < 0.001 | 0.830 (0.809–0.850), < 0.001 | 0.722 (0.696–0.749), < 0.001 |
| Number of libraries $^{vl}$ | 0.968 (0.944–0.993), 0.0110 | 1.513 (1.476–1.552), < 0.001 | 1.702 (1.640–1.766), < 0.001 |
| Robbery rate $^{vl}$ | 1.116 (1.089–1.145), < 0.001 | 0.758 (0.739–0.778), < 0.001 | 0.845 (0.817–0.874), < 0.001 |
| Theft from motor vehicle rate $^{vl}$ | 0.918 (0.896–0.941), < 0.001 | 0.747 (0.729–0.766), < 0.001 | 0.902 (0.871–0.933), < 0.001 |
| Condominium occupied private dwellings $^{vl}$ | 1.540 (1.488–1.594), < 0.001 | 0.668 (0.649–0.687), < 0.001 | 0.969 (0.935–1.004), 0.0794 |
| Non-condominium occupied private dwellings $^{vl}$ | 0.419 (0.404–0.435), < 0.001 | 1.193 (1.161–1.227), < 0.001 | 0.881 (0.846–0.917), < 0.001 |
| Religion - Christian $^{vl}$ | 1.209 (1.172–1.248), < 0.001 | 0.573 (0.555–0.591), < 0.001 | 0.756 (0.730–0.784), < 0.001 |
| Unsuitable housing $^{vl}$ | 0.726 (0.702–0.750), < 0.001 | 0.859 (0.838–0.880), < 0.001 | 0.619 (0.594–0.646), < 0.001 |

Several strong and clinically meaningful associations were observed across sensor-derived and neighborhood-level variables. For instance, greater day-to-day variability in acceleration signals was positively associated with higher SIS and RAPA scores, indicating that more dynamic movement patterns corresponded to greater social interaction and higher physical activity. Higher SIS scores reflect lower levels of social isolation and therefore greater social interaction[37]. At the neighborhood level, higher rates of theft from motor vehicles were associated with lower SIS, OHS, and RAPA scores, suggesting that exposure to less secure environments may discourage social participation and be linked to poorer physical function and reduced activity. Similarly, a higher number of community centers showed a strong positive association with SIS, highlighting the close relationship between community exposure and social interaction. Additionally, higher levels of unsuitable housing were consistently associated with lower SIS, OHS, and RAPA scores, indicating that suboptimal housing conditions may constrain social engagement, physical functioning, and activity levels.

To validate how GEOFRAIL, which integrates multimodal sensor data with neighborhood-level context, can be used to develop machine-learning models for predicting clinical outcomes, Categorical Boosting (CatBoost)[49] and Tabular Prior-Data Fitted Networks (TabPFN)[50] were trained to model representative outcomes. Models were evaluated under 5-fold cross-validation and leave-one-participant-out (LOPO) cross-validation schemes. Evaluations were conducted within a nested cross-validation framework, in which feature selection was performed using recursive feature elimination on the training data of each fold and model performance was subsequently assessed on the corresponding held-out test sets. Table 3 presents performance comparisons among models trained on sensor data, geospatial neighborhood-level data from participants' home environments, and combined sensor and geospatial neighborhood-level data.

Table 3. Performance of Categorical Boosting (CatBoost) and Tabular Prior-Data Fitted Networks (TabPFN) for regression of representative clinical outcomes, including the Social Isolation Scale (SIS), Oxford Hip Score (OHS), and Rapid Assessment of Physical Activity (RAPA). Results are reported for models trained and evaluated using multimodal sensor data, geospatial neighborhood-level data, and their combination. †Each cell reports the mean and standard deviation of mean absolute error across folds from 5-fold cross-validation and Leave-One-Participant-Out (LOPO) cross-validation schemes.

| Cross Validation | Model | Data | SIS | OHS | RAPA |
| --- | --- | --- | --- | --- | --- |
| 5-fold | CatBoost | Sensor | 1.3770 ± 0.0870† | 2.9825 ± 0.2890 | 0.2885 ± 0.0156 |
| 5-fold | CatBoost | Geo | 1.4353 ± 0.1608 | 2.4820 ± 0.1410 | 0.4071 ± 0.0432 |
| 5-fold | CatBoost | Sensor + Geo | 0.9892 ± 0.0984 | 1.6875 ± 0.1685 | 0.1660 ± 0.0335 |
| 5-fold | TabPFN | Sensor | 0.8106 ± 0.0645 | 0.6667 ± 0.0901 | 0.0803 ± 0.0397 |
| 5-fold | TabPFN | Geo | 1.5194 ± 0.1255 | 2.3122 ± 0.1317 | 0.3616 ± 0.0534 |
| 5-fold | TabPFN | Sensor + Geo | 0.6667 ± 0.0901 | 0.9555 ± 0.1174 | 0.0599 ± 0.0265 |
| LOPO | CatBoost | Sensor | 3.4651 ± 1.4412 | 7.9224 ± 5.1977 | 1.0618 ± 0.7280 |
| LOPO | CatBoost | Geo | 3.1371 ± 1.5443 | 7.1722 ± 5.5382 | 1.1093 ± 0.6657 |
| LOPO | CatBoost | Sensor + Geo | 3.3648 ± 1.6323 | 7.9838 ± 5.4920 | 1.1342 ± 0.6202 |
| LOPO | TabPFN | Sensor | 3.4035 ± 1.4547 | 9.1805 ± 5.7234 | 0.9852 ± 0.6612 |
| LOPO | TabPFN | Geo | 3.2629 ± 1.5659 | 9.2316 ± 5.1754 | 1.1319 ± 0.7221 |
| LOPO | TabPFN | Sensor + Geo | 3.3767 ± 1.7658 | 8.2234 ± 5.2110 | 1.1438 ± 0.6826 |

Across all three clinical outcomes, lower mean absolute errors were consistently observed under the 5-fold cross-validation setting, indicating more stable performance compared with Leave-One-Participant-Out (LOPO) cross-validation. The reduced performance under LOPO is expected, as recovery trajectories are highly individualized and person-specific, making generalization to entirely unseen participants more challenging. While TabPFN outperformed CatBoost under the 5-fold cross-validation setting, CatBoost showed slightly better performance under LOPO cross-validation. Importantly, although models trained on multimodal sensor data generally outperformed those using geospatial neighborhood-level data, combining sensor and neighborhood information resulted in improved performance across all 5-fold and most LOPO settings, underscoring the added predictive value of environmental context. These findings highlight the importance of incorporating neighborhood-level information for accurately monitoring and modeling recovery trajectories among community-dwelling patients whose functional and social outcomes are shaped by the environments they routinely encounter.

## Usage Notes

This section provides practical guidance on data organization, preprocessing and imputation strategies, modeling assumptions, and important considerations for reuse, including analytical extensions, prior dataset versions, and population- and context-specific limitations of the GEOFRAIL dataset.

The predictive modeling results reported in Table 3 treat the dataset in a tabular format, using daily features extracted from multimodal sensors together with neighborhood-level characteristics of participants' home locations. Researchers may extend these analyses by modeling the sensor data as longitudinal sequences across consecutive days or weeks[24], or by incorporating neighborhood-level characteristics of locations visited in addition to home environments. Given the Canadian study setting, seasonal effects may influence recovery trajectories[51], and incorporating temporal or seasonal indicators derived from timestamps may be informative for future analyses.

Neighborhood-level variables exhibit minimal missingness, with complete coverage for amenities and crime rate tables and only limited missing values in census-derived measures. For sensor data, approximately 17.6 percent of observations were missing and were imputed at the sensor level prior to feature extraction. When daily sensor recordings were missing for a participant, values were imputed using the mean of that participant's available days within the same calendar week.

An earlier subset of the data, comprising a small number of participants and lacking frailty measures and neighborhood-level contextual variables, was previously reported by Abedi et al.[24]; the present release substantially extends this work by expanding the cohort and incorporating detailed frailty assessments alongside linked neighborhood-level information.

While the dataset offers a unique combination of multimodal sensing and contextual data in a clinically defined population, its scope reflects the practical challenges of recruiting participants who meet strict inclusion criteria and sustaining long-term data collection. The number of participants enrolled over four years was relatively small; however, the volume of collected data was substantial, with rich multimodal measurements obtained longitudinally over approximately eight weeks per participant. Data collection was conducted among urban-dwelling older adults living alone in the Greater Toronto Area, and models trained on this dataset should therefore be interpreted with consideration of this specific geographic and demographic context, which may limit direct generalizability to other populations or settings.

## Data Availability

The GEOFRAIL dataset is publicly available on Zenodo[48]. It is organized into seven interconnected tables comprising demographics, sensor-derived features, clinical measures, temporal location information, amenities, crime rates, and census characteristics. The amenities, crime rate, and census tables were linked to the core dataset using external open data sources, including the Google Places API[41], the Toronto Police Service Neighborhood Crime Rates Open Data[43], and the Statistics Canada Census Profile[45].

## Acknowledgements

The authors acknowledge Dhanyasri Maddiboina, Dherya Jain, and Zara Hasan for their support throughout the data collection process, including sensor installation and removal and periodic clinical assessments.

## Author Contributions

C.H.C. and S.S.K. conceived the MAISON platform. A.A. designed and developed the MAISON platform, including front-end, hardware, and back-end cloud components. A.A. prepared and submitted the required research ethics documentation to enable data collection and dataset publication. A.A. and C.H.C. conceptualized the integration of multimodal sensor and clinical data with neighborhood-level information. A.A. conducted data preprocessing, feature extraction, neighborhood data acquisition and linkage, statistical analysis, predictive modeling, and interpretation. S.S.K validated the predictive modeling results. A.A. drafted the manuscript, which was reviewed by all authors.

## Competing Interests

The authors declare no competing interests.

## Funding

This work was supported by C.H.C. and S.S.K. funding from Centre for Aging + Brain Health Innovation (CABHI) Grants, as well as C.H.C. and S.S.K.'s Natural Sciences and Engineering Research Council of Canada (NSERC) Discovery Grants.

## Code Availability

Example scripts for linking and processing the tables in the dataset for downstream tasks, including statistical analysis and machine-learning–based predictive modeling, are provided in an accompanying GitHub repository (https://github.com/abedidev/geofrail).